\documentclass[useAMS,usenatbib]{mn2e}
\usepackage{epsf,psfrag,graphicx}
\usepackage[usenames]{color}

\newcommand{\ifm}[1]{\relax\ifmmode#1\else$\mathsurround=0pt #1$\fi}
\newcommand{\pc}{\,{\rm pc}}
\newcommand{\Mv}{\hbox{$M_{\rm vir}$}}
\newcommand{\kpc}{\,{\rm kpc}}

\newcommand{\prop}{\propto}

\def\gtrsim{\lower.5ex\hbox{$\; \buildrel > \over \sim \;$}}

\def\MN{MN}
\def\CD{ART}
\def\C14{A$_{\rm RP}$}
\def\cmmc{\,{\rm cm}^{-3}}
\def\msun{{\rm M_\odot}}

\def\fa{4.0}

\title[Streaming rates into high-$z$ galaxies]
{Distribution of streaming rates into high-redshift galaxies}

\author[Tobias Goerdt et al.]
{\parbox[t]{\textwidth}{Tobias
Goerdt$^1$\thanks{\textcolor{blue}{tobias.goerdt@univie.ac.at}}, Daniel
Ceverino$^{2, 3}$, Avishai Dekel$^4$ and Romain Teyssier$^5$}\\
\vspace*{3pt} \\
$^1$Institut f\"ur Astrophysik, T\"urkenschanzstra{\ss}e 17, Universit\"at
Wien, 1180 Wien, \"Osterreich\\
$^2$Centro de Astrobiolog{\'i}a (CSIC-INTA), Ctra de Torrej{\'o}n a Ajalvir, km
4, 28850 Torrej{\'o}n de Ardoz, Madrid, Espa\~na\\
$^3$Astro-UAM, Universidad Aut\'onoma de Madrid, Unidad Asociada CSIC, 28049
Madrid, Espa\~na\\
$^4$Racah Institute of Physics, The Hebrew University, Jerusalem 91904,
Israel\\
$^5$Institute for Computational Science, Universit\"at Z\"urich, Winterthurer
Strasse 190, 8057 Z\"urich, Schweiz \\}

\date{Draft version \today}
\pagerange{\pageref{firstpage}--\pageref{lastpage}}
\pubyear{2015}

\begin{document}

\maketitle

\label{firstpage}

\begin{abstract}
We study the accretion along streams from the cosmic web into high-redshift
massive galaxies using three sets of \textsc{amr} hydro-cosmological
simulations. We find that the streams keep a roughly constant accretion rate as
they penetrate into the halo centre. The mean accretion rate follows the mass
and redshift dependence predicted for haloes by the EPS approximation, $\dot{M}
\propto M_{\rm vir}^{1.25} \ (1 + z)^{2.5}$. The distribution of the accretion
rates can well be described by a sum of two Gaussians, the primary
corresponding to ``smooth inflow'' and the secondary to ``mergers''. The same
functional form was already found for the distributions of specific star
formation rates in observations. The mass fraction in the smooth component is
60 - 90\%, insensitive to redshift or halo mass. The simulations with strong
feedback show clear signs of reaccretion due to recycling of galactic winds.
The mean accretion rate for the mergers is a factor 2 - 3 larger than that of
the smooth component. The standard deviation of the merger accretion rate is
0.2 - 0.3 dex, showing no trend with mass or redshift. For the smooth component
it is 0.12 - 0.24 dex.
\end{abstract}


\section{Introduction}
\label{sec:intro}

The dominant idea of galaxy formation has changed recently: Decades ago it was
though \citep{rees, silk, white} that galaxies collect their baryons through
diffuse gas symmetrically falling into dark matter haloes and being
shock-heated as it hits the gas residing in them -- ’hot mode accretion’. The
mass of the halo decides if the gas will eventually settle into the galaxy. It
was shown by theoretical work and simulations \citep{fardal, bd03, keresa,
keresb, db06, ocvirk, DekelA_09a, dekel13} that galaxies at high redshift $(z
\gtrsim 2)$, acquire their baryons primarily via cold streams of relatively
dense and pristine gas with temperatures around $10^4$ K that penetrate through
the diffuse shock-heated medium -- ’cold mode accretion’. The peak of the
stream activity is around redshift 3. Theoretical work predicted a quenching of
the gas supply into high mass galaxies $(M_{\rm vir} > 10^{12}$ M$_\odot)$ at low
redshifts $(z < 2)$ \citep{db06}.

\citet{genel10} showed with $N$-body simulations that about half of the
dark-matter haloes' mass is built-up smoothly. So as the galaxies grow baryons
should also be accreted semi-continuously. Indeed \citet{DekelA_09a} showed
with hydrodynamical cosmological simulations that about two thirds of the mass
are brought in by the rather smooth gas components, that also consist of
mini-minor mergers with mass ratio smaller than 1:10. The smooth and steady
accretion through the cold streams may have been the main driver of the
formation of the massive, clumpy and star-forming discs that have been observed
at $z\sim 2$ \citep{genzel08, genel, foerster2, foerster3}. Major merger events
on the other hand may have only contributed the smaller part \citep{oscarb, cd,
cip}.

The times scales for the gas depletion of the galaxies are relatively short
compared to the Hubble time during all of cosmic history \citep{daddi,
genzel10}. So galaxies must always accrete fresh gas coming from the
intergalactic medium in order to sustain star formation over such a long time
at the observed level. The system is self-regulated: the accretion is setting
the star-formation rate independent of the amount of available gas
\citep{bouche}. The galaxy's star formation rates are fundamentally limited by
the gas inflow. Local physics like the regulation by feedback seem to be only
of secondary importance. A fact that highlights the accretion rates'
importance: the timescales for star formation rate and star formation both
crucially depend on the gas accretion rate. A robust approximation for the
average growth rate of halo virial mass $M_{\rm vir}$ has been derived
\citep{NeisteinE_06a, neistein08} using the EPS \citep{lacey} theory of
cosmological clustering into spherical haloes in virial equilibrium.

Attempts to prove the cold accretion stream paradigm observationally are
ongoing: The characteristics of Ly$\alpha$ emission produced by the cold gas
streams via the release of gravitational energy\footnote{The gas remains at
constant velocity as it flows down the potential gradient towards the halo
centre \citep{mich4}.} have been predicted using cosmological hydrodynamical
\textsc{amr} simulations \citep{mich}. The simulated Ly$\alpha$-blobs (LABs)
can reproduce many of the features of the observed LABs. They can therefore be
interpreted as direct observational detections of cold stream accretion.
Likelihoods of observing streams in absorption have been theoretically
predicted \citep{mich2}. A planar subgroup of satellites has been found to
exist in the Andromeda galaxy \citep[M31;][]{ibata}. It comprises about half of
the population. This vast thin disk of satellites is a natural result of cold
stream accretion \citep{mich3} and can therefore also be treated as
observational evidence for the existence of cold streams.

Observationally two main modes of star formation are known to control the growth
of galaxies: a relatively steady one in disk-like galaxies and a starburst mode
which is generally interpreted as driven by merging. \citet{rodighiero}
quantified the relative contribution of the two modes in the redshift interval
$1.5 < z < 2.5$ using PACS/{\it Herschel} observations over the whole COSMOS
and GOODS-South fields, in conjunction with previous optical/near-infrared data.
\citet{sargent} used their data to predict the shape of the infrared luminosity
function at redshifts $z \le 2$ by introducing a double-Gaussian decomposition
of the specific star formation rate distribution at fixed stellar mass into a
contribution (assumed redshift- and mass-invariant) from main-sequence and
starburst activity.

The characteristics of baryonic accretion on to galaxies have been analysed
with the help of cosmological hydrodynamic simulations in various works. Key
results are that gas accretion is mostly smooth, with mergers only becoming
important for groups and clusters. The specific rate of the gas accretion on to
haloes is only weakly dependent on the halo mass \citep{freke2}. The accreted
gas is bimodal with a temperature division at 10$^5$ K. Cold-mode accretion
dominates inflows at early times and declines below $z \sim 2$. Hot-mode
accretion peaks near $z = 1 - 2$ and declines gradually. Cold-mode accretion
can fuel immediate star formation, while hot-mode accretion preferentially
builds a large, hot gas reservoir in the halo \citep{woods}. Another third
distinct accretion mode along with the 'cold' and 'hot' modes has been
identified \citep{oppenheimer}, called 'recycled wind mode', in which the
accretion comes from material that was previously ejected from a galaxy.
Galaxies in substantially overdense environments grow predominantly by a smooth
accretion from cosmological filaments which dominates the mass input from
mergers \citep{emilio}. The relationship between stellar mass, metallicity, and
star formation rate can be explained by metal-poor gas accretion
\citep{almeida}.

\begin{figure*}
\begin{center}
\psfrag{$1e1$}[Bl][Bl][1][0]{$10$}
\psfrag{$1e2$}[Bl][Bl][1][0]{$100$}
\psfrag{$1e3$}[Bl][Bl][1][0]{$1000$}
\psfrag{$1e4$}[Bl][Bl][1][0]{$10^4$}
\psfrag{dM/dt /Om [MO / yr / sr]}[B][B][1][0]{$\dot{M} \ \Omega^{-1}$
[M$_\odot$ yr$^{-1}$ sr$^{-1}$]}
\includegraphics[width=14.00cm]{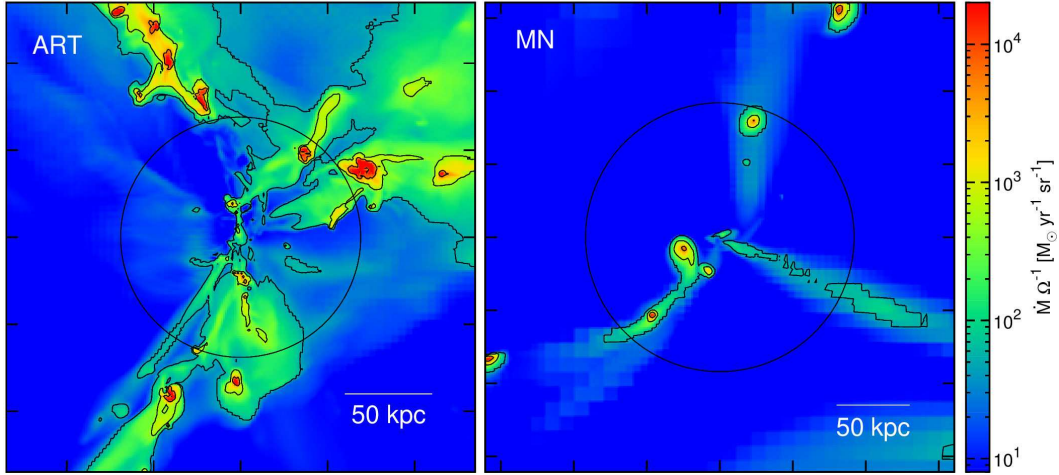}
\end{center}
\caption{Simulated galaxies from two different suites of simulations ({\MN} and
\CD). Shown in colour coding is the peak gas inflow along the line of sight.
Inflow is here defined as the radial gas inflow per unit solid angle. The
contours indicate gas influxes of $\dot{M} \ \Omega^{-1} = 50$, 500 and $5000$
M$_\odot$ yr$^{-1}$ sr$^{-1}$, respectively. The circles refer to the virial
radii. Left: one of the {\CD} galaxies (resolution $70\pc$) at $z=2.3$, with
$\Mv = 3.5 \times 10^{11}\msun$. Right: a typical {\MN} galaxy (resolution
$1\kpc$) at $z=2.5$, with $\Mv = 10^{12}\msun$. The inflow is dominated in both
cases by three cold narrow streams that are partly clumpy.}
\label{fig:denmap}
\end{figure*}

The properties of cold- and hot-mode gas, are clearly distinguishable in the
outer parts of massive haloes. The cold-mode gas is confined to clumpy
filaments that are approximately in pressure equilibrium with the hot-mode gas.
Cold- mode gas typically has a much lower metallicity and is much more likely
to be infalling \citep{freke1}. Stream morphologies become increasingly complex
at higher resolution, with large coherent flows revealing density and
temperature structure at progressively smaller scales \citep{dylan}.

The cold gas accretion rate is not an universal factor of the dark matter
accretion rate. Galactic winds can cause star formation rates to deviate
significantly from the external gas accretion rates, both via gas ejection and
reaccretion. Cold accretion is broadly consistent with driving the bulk of the
highly star-forming galaxies observed at $z \sim 2$ \citep{faucher}. The key
physical variables of galactic cold inflows substantially decrease with
decreasing redshift, namely: the number of streams, the average inflow rate per
stream as well as the mean gas density in the streams. The stream angular
momentum on the other hand is found to increase with decreasing redshift
\citep{cen}.

The average radial velocities of the inflowing material decrease with
decreasing radius \citep[roughly a decrease of 65\% between 1 $r_{\rm vir}$ and
the centre of the host;][their figures 4 and 6, bottom panels]{wetzel}. The
velocity profile of the gas flowing along the streams into a galaxy's halo in
the form of cold streams is, contrary to what might be expected, roughly
constant with radius instead of free-falling \citep{mich4}.

In a forthcoming companion paper \citep{mich5} we are going to address the role
of mergers versus smooth flows by analysing the clumpiness of the gas streams.
We will evaluate each clump mass and estimate a mass ratio for the expected
merger. Finally we are going to look at the distribution of constituents
amongst the clumps.

In this paper we look at the amount of inflow -- the mass accretion rate -- as
a function of radius, mass and redshift for the three constituents gas, stars
and dark matter. The paper is organised as follows: in section \ref{sec:sim} we
present the three suites of simulations used for the analysis. In section
\ref{sec:raw} we show individual and averaged inflow curves. In section
\ref{sec:distri} we look at the distributions of the amount of inflow, i.e. at
the deviations from the means presented in section \ref{sec:raw}. In section
\ref{sec:conc} we draw our conclusions.

\section{simulations}
\label{sec:sim}

Galaxy snapshots of three different suites of cosmological simulations based on
Eulerian \textsc{amr} (Adaptive Mesh Refinement) hydrodynamics are analysed for
this paper. The three suites are the Adaptive Refinement Tree \citep[hereafter
\CD;][]{cd, c12, dekel13, cip}, the {\C14} \citep{c14, adi} as well as the
Horizon-MareNostrum \citep[hereafter \MN;][]{ocvirk} simulation. The {\CD} and
the {\C14} suites of simulations both consist of several simulations zooming in
with a maximum resolution of $15-70\pc$ at $z = 2$ on individual galaxies that
reside in dark-matter haloes of masses $(0.07 - 1.98) \times 10^{12}\msun$ at
$z = 2.3$. The {\MN} simulation contains hundreds of massive galaxies in a
cosmological box of side 71 Mpc with a maximum resolution of $\simeq 1\kpc$.

To show the morphology of the gas streams we present in figure \ref{fig:denmap}
gas influx maps of two sample galaxies from the {\MN} and the {\CD} suite. The
panels demonstrate the dominance of typically three, narrow cold streams, which
come from well outside the virial radius along the dark-matter filaments of the
cosmic web, and penetrate into the discs at the halo centres. The streams are
partly clumpy and partly smooth, even in the simulations with higher
resolution.

\subsection{High-resolution {\CD} simulations}

The {\CD} simulations were run with the \textsc{amr} code \textsc{art}
\citep[Adaptive Refinement Tree;][]{kkk,andrey} with a spatial resolution better
than $70 \pc$ in physical units. It incorporates the relevant physical
processes for galaxy formation that are: gas cooling, photoionisation heating,
star formation, metal enrichment and stellar feedback \citep{cak}. Cooling
rates were computed for the given gas density, temperature, metallicity, and UV
background based on \textsc{cloudy} \citep{ferland}. Cooling is assumed at the
centre of a cloud of thickness 1 kpc \citep{ceverino, rk}. Metallicity
dependent, metal-line cooling is included, assuming a relative abundance of
elements equal to the solar composition. The code implements a ``constant"
feedback model, in which the combined energy from stellar winds and supernova
explosions is released as a constant heating rate over 40 Myr (the typical age
of the lightest star that can still explode in a type-II supernova).
Photo-heating is also taken into account self-consistently with radiative
cooling. A uniform UV background based on the \citet{haardt} model is assumed.
Local sources are ignored. In order to mimic the self-shielding of dense,
galactic neutral hydrogen from the cosmological UV background, the simulation
assumes for the gas at total densities above $n=0.1$ cm$^{-3}$ a substantially
suppressed UV background ($5.9 \times 10^{26} \ {\rm erg} \ {\rm s}^{-1} \ {\rm
cm}^{-2} \ {\rm Hz}^{-1}$, the value of the pre-reionisation UV background at
$z = 8$). The self-shielding threshold of $n = 0.1$ cm$^{-3}$ is set by
radiative transfer simulations of ionising photons coming from the cosmological
background as well as from the galaxy itself \citep[see appendix A2
of][]{michele}.

The \textsc{art} code has a unique feature for the purpose of simulating the
detailed structure of the streams. It allows gas cooling to well below $10^4$K.
This enables high densities in pressure equilibrium with the hotter and more
dilute medium. A non-thermal pressure floor has been implemented to ensure that
the Jeans length is resolved by at least seven resolution elements and thus
prevent artificial fragmentation on the smallest grid scale \citep{truelove,
rk, cd}. It is effective in the dense $(n > 10$ cm$^{-3})$ and cold $(T<10^4
K)$ regions inside galactic disks.

The equation of state remains unchanged at all densities. Stars form in cells
where the gas temperature is below $10^4$K and the gas density is above a
threshold of $n = 1\cmmc$ according to a stochastic model that is consistent
with the \citet{kennicutt} law. In the creation of a single stellar particle, a
constant fraction (1/3) of the gas mass within a cell is converted into stellar
mass \citep[see appendix of][]{cak}. The ISM is enriched by metals from
supernovae type II and type Ia. Metals are released from each star particle by
SNII at a constant rate for 40 Myr after its birth. A \citet{miller} initial
mass function is assumed which is matching the results of \citet{woosley}. The
metal ejection by SNIa assumes an exponentially declining SNIa rate from a
maximum at 1 Gyr. The code treats the advection of metals self-consistently and
it distinguishes between SNII and SNIa ejecta \citep{ceverino}.

The initial conditions for the {\CD} simulations were created using
low-resolution cosmological $N$-body simulations in comoving boxes of side
29 - 114\,Mpc. Its cosmological parameters were motivated by WMAP5
\citep{WMAP5}. The values are: $\Omega_{\rm m} = 0.27$, $\Omega_\Lambda = 0.73$,
$\Omega_{\rm b} = 0.045$, $h = 0.7$ and $\sigma_8 = 0.82$. We selected 31 haloes
of $\Mv \simeq 10^{12}\msun$ at $z = 1.0$. For each halo, a concentric sphere
of radius twice the virial radius was identified for re-simulation with high
resolution. Gas was added to the box following the dark matter distribution
with a fraction $f_{\rm b} = 0.15$. The whole box was then re-simulated, with
refined resolution in the selected volume about the respective galaxy. The dark
matter particle mass is $5.5 \times 10^5 \msun$, the minimum star particle mass
is $10^4 \msun$, the smallest cell size is 35 pc (physical units) at $z = 2$.

\subsubsection{Simulations including radiation pressure (\C14)}

The {\C14} suite of simulations \citep{c14, adi} is a further development of
last subsection's {\CD} suite: Apart from the features already presented there,
it also includes the effects of radiation pressure by massive stars. The
radiation pressure was modelled as a non-thermal pressure that acts only in
dense and optically thick star-forming regions in a way that the ionising
radiation injects momentum around massive stars, pressurising star-forming
regions \citep[][their appendix B]{oscarc}. The {\C14} suite uses a stronger
feedback that brings the stellar-to-halo mass ratio closer to the estimates by
abundance matching \citep{moster, behroozi}. See \citet{c14} and \citet{moody}.
The adaptive comoving mesh has been refined in the dense regions to cells of
minimum size between $17 - 35$ pc in physical units. The DM particle mass is
$8.3 \times 10^4$ M$_\odot$. The particles representing star clusters have a
minimum mass of $10^3$ M$_\odot$, similar to the stellar mass of an Orion-like
star cluster. The initial conditions for the {\C14} simulations were created
using low-resolution cosmological $N$-body simulations in comoving boxes of
side 14 - 57\,Mpc. We selected 34 haloes of $\Mv = 1.18 - 14.7 \times
10^{11}\msun$ at $z = 1.0$.

\subsection{RAMSES Horizon-MareNostrum simulation}

The {\MN} simulation uses the \textsc{amr} code \textsc{ramses}
\citep{teyssier}. The spatial resolution is $\sim 1\,$kpc in physical units. UV
heating is included assuming the \citet{haardt} background model, as in the
{\CD} simulation. The code incorporates a simple model of supernovae feedback
and metal enrichment using the implementation described in \citet{dubois}. The
cooling rates are calculated assuming ionisation equilibrium for H and He,
including both collisional ionisation and photoionisation \citep{katz}. Metal
cooling is also included using tabulated \textsc{cloudy} rates, and is assumed
proportional to the metallicity, relative to the \citet{grevesse} solar
abundances. Unlike in the {\CD} simulation, no cooling below $T<10^4$K is
computed, and no self-shielding of the UV flux is assumed.

For high-density regions, the \textsc{ramses} code considers a polytropic
equation of state with $\gamma_0 = 5/3$ to model the complex, multi-phase and
turbulent structure of the inter-stellar medium (ISM) \citep{yepes, sh} in a
simplified form \citep[see][]{joop, dubois}. The ISM is defined as gas with
hydrogen density greater than $n_{\rm H} = 0.1\cmmc$, one order of magnitude
lower than in the {\CD} simulation. Star formation has been included, for ISM
gas only, by spawning star particles at a rate consistent with the
\citet{kennicutt} law derived from local observations of star forming galaxies. 

\begin{table}
\begin{center}
\setlength{\arrayrulewidth}{0.5mm}
\begin{tabular}{rcccr}
\hline
label & suite & $M_{\rm vir}$ [10$^{12}$ M$_\odot$] & $z$ & $N_{\rm gal}$ \\
\hline
10$^{13}$            & {\MN} & 10.47 $\pm$ 0.56   & 1.57 & 12 \\
10$^{13}$            & {\MN} & 10.49 $\pm$ 0.93   & 2.46 & 12 \\
$5 \times 10^{12}$   & {\MN} & 5.00 $\pm$ 0.045 & 1.57 & 12 \\
$5 \times 10^{12}$   & {\MN} & 5.48 $\pm$ 0.26   & 2.46 & 11 \\
10$^{12}$            & {\MN} & 1.03 $\pm$ 0.003 & 1.57 &  8 \\
10$^{12}$            & {\MN} & 1.01 $\pm$ 0.004 & 2.46 & 12 \\
10$^{12}$            & {\MN} & 1.03 $\pm$ 0.006 & 4.01 &  9 \\
10$^{11}$            & {\MN} & 0.099 $\pm$ 0.000 & 1.57 & 12 \\
10$^{11}$            & {\MN} & 0.099 $\pm$ 0.000 & 2.46 &  7 \\
10$^{11}$            & {\MN} & 0.099 $\pm$ 0.000 & 4.01 & 12 \\
$1.9 \times 10^{12}$ & {\CD} & 1.907 $\pm$ 0.217 & 1.14 $\pm$ 0.02  & 34 \\
$1.3 \times 10^{12}$ & {\CD} & 1.286 $\pm$ 0.093 & 1.60 $\pm$ 0.02  & 73 \\
$8.6 \times 10^{11}$ & {\CD} & 0.863 $\pm$ 0.046 & 2.25 $\pm$ 0.02  & 109 \\
$3.9 \times 10^{11}$ & {\CD} & 0.391 $\pm$ 0.034 & 3.40 $\pm$ 0.04  & 119 \\
$7.1 \times 10^{11}$ & {\C14} & 0.707 $\pm$ 0.055 & 1.14 $\pm$ 0.02  & 41 \\
$6.7 \times 10^{11}$ & {\C14} & 0.672 $\pm$ 0.049 & 1.58 $\pm$ 0.02  & 47 \\
$4.9 \times 10^{11}$ & {\C14} & 0.491 $\pm$ 0.028 & 2.21 $\pm$ 0.03  & 57 \\
$2.9 \times 10^{11}$ & {\C14} & 0.290 $\pm$ 0.019 & 3.27 $\pm$ 0.05  & 62 \\
$2.6 \times 10^{11}$ & {\C14} & 0.260 $\pm$ 0.010 & 1.13 $\pm$ 0.02  & 40 \\
$2.4 \times 10^{11}$ & {\C14} & 0.242 $\pm$ 0.010 & 1.59 $\pm$ 0.02  & 50 \\
$1.6 \times 10^{11}$ & {\C14} & 0.160 $\pm$ 0.007 & 2.29 $\pm$ 0.03  & 60 \\
$7.3 \times 10^{10}$ & {\C14} & 0.073 $\pm$ 0.004 & 3.53 $\pm$ 0.05  & 66 \\
\hline
\end{tabular}
\end{center}
\caption{The various bins of galaxies for our analyses. They are used
throughout the whole paper. 'Label' is the tag the corresponding bin is
labelled by in our figures. It is a mass close to the ensemble's actual mean
virial mass. 'Suite' denotes the suite of simulations the bin stems from.
$M_{\rm vir}$ is the mean virial mass of the bin together with its standard
deviation. $z$ is the mean redshift of the ensemble together with its standard
deviation. Galaxies from different {\MN} snapshots are not combined therefore
the redshift's standard deviation is always zero for all {\MN} bins.
$N_{\rm gal}$ gives the total number of galaxies in the given bin.}
\label{tab:bins}
\end{table}

The {\MN} simulation implemented a pressure floor in order to prevent
artificial fragmentation, by keeping the Jeans lengthscale, $\lambda_{\rm J}
\prop T n^{-2/3}$, larger than the size of four grid cells everywhere.
At every position where $n>0.1\cmmc$, a density dependent temperature floor was
imposed. It mimics the average thermal and turbulent pressure of the multiphase
ISM \citep{sh, dvs}. The gas is allowed to heat up above this temperature floor
and cool back again. The temperature floor follows a polytropic equation of
state with $T_{\rm floor}=T_0 (n/n_0)^{\gamma_0-1}$, where $T_0=10^4$ K and $n_0=$
0.1 cm$^{-3}$. The resulting pressure floor is given by $P_{\rm floor} =
n_{\rm H} \ k_{\rm B} \ T_{\rm floor}$.

For each stellar population, 10\% of the mass is assumed to turn into
supernovae type II after 10 Myr, where the energy and metals are released in a
single impulse. For each supernova, 10\% of the ejected mass is assumed to be
pure metals, with the remaining 90\% keeping the metallicity of the star at
birth. SNIa feedback has not been considered. The initial conditions of the
{\MN} simulation were constructed assuming a $\Lambda$CDM universe with
$\Omega_{\rm M} = 0.3$, $\Omega_\Lambda = 0.7$, $\Omega_{\rm b} = 0.045$, $h=0.7$
and $\sigma_8 = 0.9$ in a periodic box of 71 Mpc. The adaptive-resolution rules
in this simulation were the same everywhere, with no zoom-in resimulation of
individual galaxies. The dark matter particle mass is $1.16 \times 10^7 \msun$,
the star particle mass is $2.05 \times 10^6 \msun$, the smallest cell size is
$1.09$ kpc physical, and the force softening length is 1.65 kpc.

We will always show averaged results for an ensemble of galaxies having very
similar masses and redshifts. In table \ref{tab:bins} we show a summary of the
various bins of galaxies we use. Since we have better statistics for the {\MN}
simulation we were able to bin galaxies from a narrower mass range therefore
the standard deviations of the mean mass of a {\MN} bin is usually much
smaller. To partly compensate for that we combine {\CD} galaxies from adjacent
redshifts increasing the statistics but introducing a standard deviation into
the mean redshift of the bin.

\section{Amount of inflow}
\label{sec:raw}

First the average inflow is computed as a function of radius from the
simulations. To do so the amount of mass is measured that crosses within a
small time $\Delta t$ a spherical shell of radius $r$ that is centred around
any given host galaxy. This is done independently for gas, stars and dark
matter, which are the three different constituents in the simulations. To get
an inflow rate, we divide the mass that is crossing through a given shell by
the time $\Delta t$ taken. The reader should note that we do include the
expansion due to the Hubble flow in this calculation. As with most cosmological
codes only the peculiar velocities are written out into the snapshots onto hard
disk. We then add the velocity of the Hubble expansion as measured from the
centre of the respective host halo by hand. The crucial difference between
stars and dark matter particles on the one hand and gas on the other hand is
that stars and dark matter are collisionless and therefore virialised within
the virial radius $r_{\rm vir}$, whereas gas on the other hand is neither. In
case of the gas inflow we chose to only take into account inwards radial
velocity cells in order to get the cleanest estimate for the amount of the
inflow. Whereas for stars and dark matter we have to take into account all the
material (i.e. only the net influx). If one takes into account only the
inflowing stellar particles, the difference would not be very big, the stellar
inflow rates would only be $\sim 25\%$ bigger. The stars follow the gas very
precisely. For the dark matter particles on the other hand the differences are
tremendous: The inflow rates increase by more than a factor of two. It is worth
stressing that we do not use any information about the temperature: so hot
inflowing gas is taken into account, too. The hot gas however contributes very
little to the inflow, since the vast majority of the mass of the material is in
the cold phase.

\begin{figure}
\begin{center}
\psfrag{$0.2$}[B][B][1][0]{0.2}
\psfrag{$0.4$}[B][B][1][0]{0.4}
\psfrag{$0.6$}[B][B][1][0]{0.6}
\psfrag{$0.8$}[B][B][1][0]{0.8}
\psfrag{$1$}[B][B][1][0]{1}
\psfrag{$1.2$}[B][B][1][0]{1.2}
\psfrag{$1.4$}[B][B][1][0]{1.4}
\psfrag{$1.6$}[B][B][1][0]{1.6}
\psfrag{$10$}[B][B][1][0]{10}
\psfrag{$70$}[B][B][1][0]{70}
\psfrag{$80$}[B][B][1][0]{80}
\psfrag{$90$}[B][B][1][0]{90}
\psfrag{$100$}[B][B][1][0]{100}
\psfrag{$200$}[B][B][1][0]{200}
\psfrag{$300$}[B][B][1][0]{300}
\psfrag{$400$}[B][B][1][0]{400}
\psfrag{$500$}[B][B][1][0]{500}
\psfrag{$600$}[B][B][1][0]{600}
\psfrag{$700$}[B][B][1][0]{700}
\psfrag{$800$}[B][B][1][0]{800}
\psfrag{$900$}[B][B][1][0]{900}
\psfrag{r [rvir]}[B][B][1][0]{$r$ [$r_{\rm vir}$]}
\psfrag{Minflo [MO/yr]}[B][B][1][0]{$\dot M$ [M$_\odot$ yr$^{-1}$]}
\psfrag{MN: 1e12 MO, z = 2.46}[Br][Br][1][0]{\textcolor{blue}{{\MN}:
$M_{\rm vir}=10^{12}\,$M$_\odot$, $z = 2.46$}}
\psfrag{CDB: 8.6e11 MO, z = 2.25}[Br][Br][1][0]{\textcolor{red}{{\CD}:
$M_{\rm vir}=8.6 \times 10^{11}\,$M$_\odot$, $z = 2.25$}}
\psfrag{ARP: 4.9e11 MO, z = 2.21}[Br][Br][1][0]{\textcolor{green}{{\C14}:
$M_{\rm vir}=4.9 \times 10^{11}\,$M$_\odot$, $z = 2.21$}}
\psfrag{gas}[Bl][Bl][1][0]{gas}
\psfrag{stars}[Bl][Bl][1][0]{stars}
\psfrag{DM}[Bl][Bl][1][0]{DM}
\includegraphics[width=6.88cm]{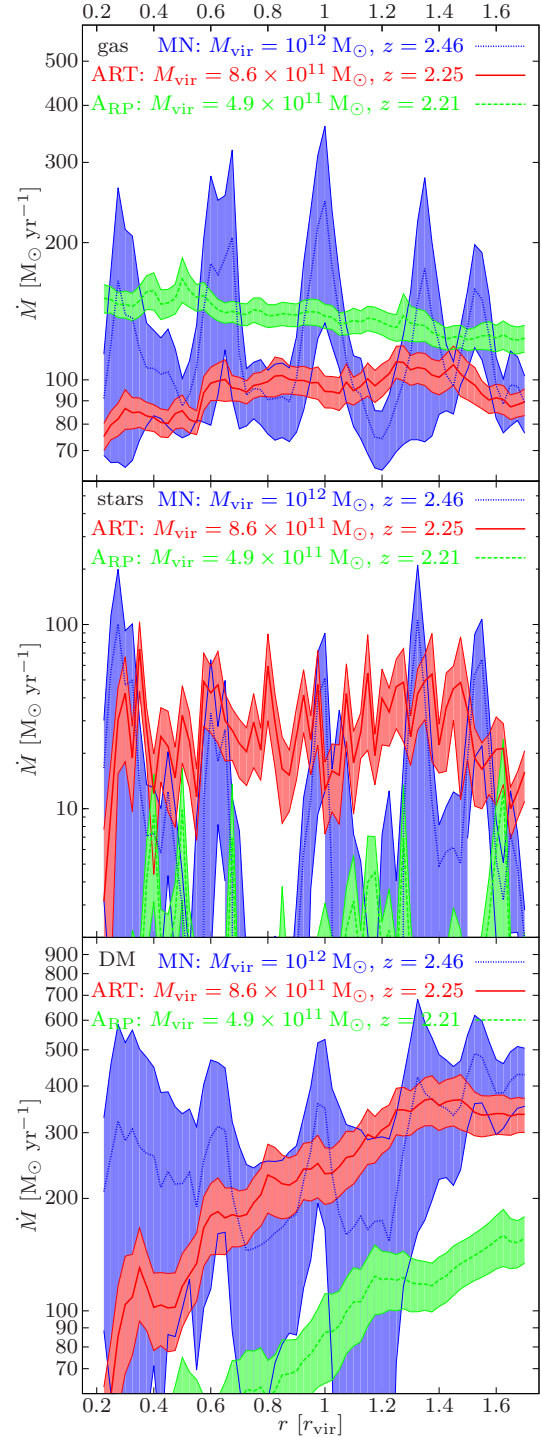}
\end{center}
\caption{The average inflow as a function of radius together with its
1$ - \sigma$ standard deviations. The {\MN} (dotted blue), the {\CD} (solid
red) and the {\C14} (dashed green) simulations are compared. In gas (top panel)
an average ``ground'' inflow is seen in gas which is fairly constant over the
different radii. On top of this smooth ground inflow there are distinct peaks
in the inflow, corresponding to bigger clumps. In the stellar distribution
(middle panel) signals of inflow are only seen at those gas peaks. Stars seem
to form only within those gas clumps. The stellar inflow rates are noisy but
roughly constant. In dark matter (bottom panel) the inflow increases with
radius. The gas inflow of the {\C14} simulation is fairly high due to its
recycling processes.}
\label{fig:avinflvsradcombi}
\end{figure}

We also want to briefly look at the inflow rates of all cells to compare those
to the ones from our choice of taking into account only the cells which have an
inwards radial velocity. The results vary for the different suites of
simulations: for the {\MN} suite, the one with the weakest feedback, there is
hardly any difference: the inflow rates are still constant at roughly the same
level. For the {\CD} simulations which has stronger feedback implemented, the
average inflow rate drops by a factor of two, when also taking into account
cells with outflowing velocities. Also we see that the inflow rates are smaller
towards the centre of the host halo. An effect already described by
\citet[][when taking into account only inflowing gas, they find approximately
constant accretion rates as well]{freke1}. For the {\C14} suite of simulations
which has by far the strongest feedback the average inflow rate drops by a
factor of ten, but for this suite the inflow rates are higher towards the
centre of the host halo. We see an inverted \citet{freke1} effect. The
differences between taking outflowing gas cells into account or not might seem
large, the reader however should note that we are mainly interested in the
inflowing material and not so much the outflowing material, so concentrating on
the inflowing cells only is the right choice.

We compute inflow rates for the galaxies. For reducing the statistical noise
that is present in the inflow of a single galaxy we stack the amount of inflow
of all available galaxies having similar redshifts and masses, from all three
suites of simulations galaxies that are available at a variety of different
halo masses and redshifts. Since the {\MN} is a fully cosmological simulation
it is possible to compile bins of galaxies spanning several orders of magnitude
in mass for various redshifts. The {\CD} suite consists only of resimulated
galaxies at comparable masses and therefore there is at any given redshift only
one mass bin available which evolves in mass with cosmic time. The situation of
the {\C14} suite is similar, however for this suite two different mass bins are
available at the given redshifts, but those masses also evolve with cosmic
time. The average halo masses and redshifts of the chosen bins are summarised
in table \ref{tab:bins}.

In figure \ref{fig:avinflvsradcombi} the average inflow as a function of radius
for galaxies at $z = 2.2 - 2.5$ with $M_{\rm vir} = 5 - 10 \times 10^{11}$
M$_\odot$ of all three {\MN}, {\CD} and {\C14} are shown. The galaxies have an
average ``ground'' inflow in gas of about 100 M$_\odot$ yr$^{-1}$ which is fairly
constant over the different radii. This is the smooth component of the inflow,
the cold streams. On top of this smooth ground inflow there are a couple of
distinct peaks. These peaks can still be seen in the averaged data, especially
if the number of galaxies in a certain bin is small (i.e. the {\MN}
simulation). Those peaks correspond to bigger clumps which are the merger
events. We use the term ``merger'' to describe any major or minor merger of
mass ratio $\mu_{\rm m} \ge 0.1$, as distinct from ``smooth flows'', which
include ``mini-minor'' mergers with mass ratios $\mu_{\rm m} < 0.1$. We will
refer to this bimodality of the modes of inflow, smooth accretion versus clumps
or merger events, throughout the whole paper.

\begin{figure*}
\begin{center}
\psfrag{MN}[Br][Br][1][0]{\MN}
\psfrag{CDB}[Br][Br][1][0]{\CD}
\psfrag{ARP}[Br][Br][1][0]{\C14}
\psfrag{ND08}[Br][Br][1][0]{ND08}
\psfrag{gas, average}[Bl][Bl][1][0]{gas, average}
\psfrag{$2$}[B][B][1][0]{$2$}
\psfrag{$2.5$}[B][B][1][0]{$2.5$}
\psfrag{$3$}[B][B][1][0]{$3$}
\psfrag{$3.5$}[B][B][1][0]{$3.5$}
\psfrag{$4$}[B][B][1][0]{$4$}
\psfrag{$4.5$}[B][B][1][0]{$4.5$}
\psfrag{$5$}[B][B][1][0]{$5$}
\psfrag{$10$}[B][B][1][0]{$10$}
\psfrag{$100$}[B][B][1][0]{$100$}
\psfrag{$1000$}[B][B][1][0]{$1000$}
\psfrag{$1e11$}[B][B][1][0]{$10^{11}$}
\psfrag{$1e12$}[B][B][1][0]{$10^{12}$}
\psfrag{$1e13$}[B][B][1][0]{$10^{13}$}
\psfrag{scaled to z = 2.46}[Bl][Bl][1][0]{scaled to $z = 2.46$}
\psfrag{scaled to M = 1e12 MO}[Bl][Bl][1][0]{scaled to $M_{\rm vir} = 10^{12}$
M$_\odot$}
\psfrag{Minflow [MO/yr]}[B][B][1][0]{$\langle \dot{M} \rangle$ [M$_\odot$
yr$^{-1}$]}
\psfrag{Mvir [MO]}[Bl][Bl][1][0]{$M_{\rm vir}$ [M$_\odot$]}
\psfrag{z + 1}[B][B][1][0]{$z + 1$}
\includegraphics[width=16.00cm]{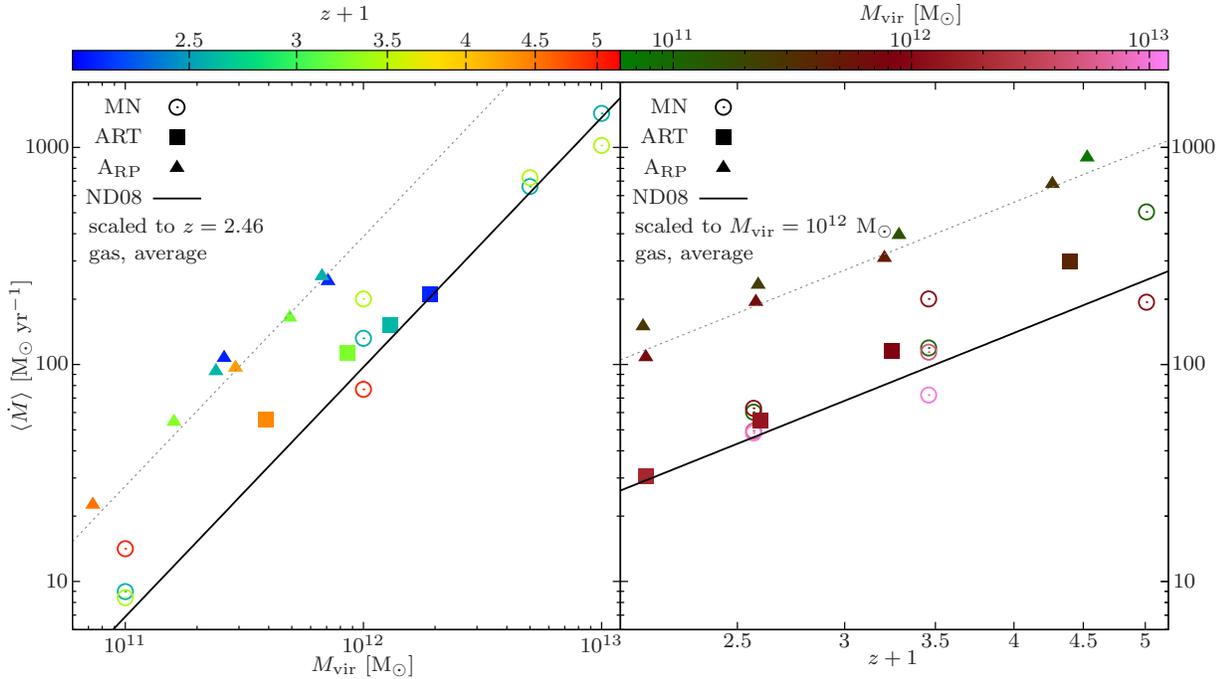}
\end{center}
\caption{The amount of inflow $M_{\rm inflow}$ of the gas as a function of halo
mass $M_{\rm vir}$ (left panel) or as a function of redshift (right panel).
The {\CD} data is plotted as filled squares, the {\MN} data as open circles and
the {\C14} data as filled triangles. The data points have been rescaled
according to $\langle \dot M \rangle \propto [1 + z]^{2.5}$ (left panel) or
$\langle \dot M \rangle \propto [M_{\rm vir}]^{1.15}$ (right panel) in order to
match $z = 2.46$ (left panel) or $M_{\rm vir} = 10^{12} \ {\rm M}_\odot$ (right
panel). The original redshift or host halo mass values are given in colour
coding. The {\C14} values are higher than the other two simulations, but they
also follow clearly the same slopes (indicated by the dashed grey line). This
is due to the fact that the {\C14} simulations have more re-accretion due to
recycling. One should correct the gas inflow in those simulations by a factor
of $f_{\rm rec}(\dot{M}) = \fa$ due to this effect. The solid black lines are the
predictions from the \citet{neistein08} model (equation \ref{eqn:neistein}).
The agreement between the data and the model is truly remarkable (Taking
$f_{\rm rec}(\dot{M})$ into account for {\C14}).}
\label{fig:Minfl}
\end{figure*}

The {\MN} simulation is in general more noisy which can be attributed to its
lower resolution. The {\C14} simulation has a lower stellar and dark matter
infall because of its much lower virial mass. The gas infall of the {\C14}
simulation on the other hand is higher than that of the {\CD} or {\MN}
simulation, which is due to the fact that the {\C14} simulation has stronger
feedback than the {\CD} suite by design. Additionally to all the feedback, the
{\CD} suite of simulation already has, the {\C14} simulation also includes
feedback by radiation pressure. \citet{c14} compared the effects of the
feedback models used in {\CD} and {\C14} suites of simulations. SFR and stellar
masses were lower by a factor $\sim 3$ with the stronger feedback. \citet{adi}
computed the outflow rates (their figures 2 and 3) of the {\C14} suite and they
found outflow rates similar or even higher than the SFR. Therefore the {\C14}
suite has more massive outflows and part of the outflowing material can rain
down back to the central galaxy. An effect usually coined ``recycling''
\citep{oppenheimer}. As we will see later in figure \ref{fig:Minfl} the gas
inflow in the {\C14} suite of simulations is enhanced due to recycling compared
to the {\MN} or the {\CD} simulations. The inflow of the {\C14} simulations is
higher by a factor $f_{\rm rec}(\dot{M}) = \fa$ compared to the {\MN} or the
{\CD} simulations due to this effect. This factor for the inflow stays constant
over the whole redshift range and also over the whole host halo mass range
considered. We will also see a similar effect of recycling later in the paper
during the fitting procedure of the distributions.

Otherwise all three suites behave similar: The gas inflow is roughly constant
at around 100 M$_\odot$ yr$^{-1}$. \citet{teklu} reports a similar behaviour in
\textsc{sph} simulations. She finds a total gas inflow $\dot{M} =
90$\,M$_\odot$\,yr$^{-1}$ for a galaxy with $M_{\rm vir} = 1.5 \times
10^{12}$\,M$_\odot$ at $z = 2.33$. Coming back to our own analysis and figure
\ref{fig:avinflvsradcombi}, we see that the {\MN} simulation seems to have a
marginally higher averaged gas inflow than the {\CD} which can be accounted for
by the 16\% more mass the average {\MN} galaxy has. The gas inflow of the
{\C14} simulation in turn is even higher, which can be accounted for by either
recycling or more efficient cooling from more efficient metal enrichment.

In {\MN} we see that there is less stellar infall, but more gas infall by
roughly a factor of $\sim 2$ compared to the {\CD} simulations. In {\C14} there
is even less stellar infall and even more gas infall. The stellar inflow
behaviour in general is extremely noisy. Its inflow rates oscillate between a
few and up to about 80 M$_\odot$ yr$^{-1}$ for both, with the {\CD} and the
{\C14} being more constant compared to {\MN}, which is due to the fact that the
bins of the {\CD} and the {\C14} simulations consists of considerably more
galaxies than the bins of the {\MN} (see table \ref{tab:bins}). Any peaks in
the inflow profile (e.g. due to merger events) are much more smoothed out by
simply the averaging process. It is noteworthy that in the stellar distribution
there is only signal of inflow at the positions of the very peaks of the gas
inflow, indicating that the smooth stream inflow component is fairly depleted
from stars. The stars seem to flow into the central galaxy in the form of
bigger clumps only. It is in this sense that the stars seem to follow the
positions of the gas as we will refer to later. The lines for the dark matter
inflow behave similar at larger radii. They exhibit slightly decreasing infall
between 400 M$_\odot$ yr$^{-1}$ at 1.7 $r_{\rm vir}$ down to 200 M$_\odot$
yr$^{-1}$ at 0.7 $r_{\rm vir}$. Inside of $\sim 0.7$ $r_{\rm vir}$ the three
suites differ: {\MN} stays constant although the scatter is larger, whereas
{\CD} and {\C14} continue to decrease. The {\CD} decreases down to 50 M$_\odot$
yr$^{-1}$ at 0.2 $r_{\rm vir}$ and the {\C14} down to 50 M$_\odot$ yr$^{-1}$
already at 0.8 $r_{\rm vir}$.

The averaged amount of inflow into the galaxies as a function of radius is by
and large constant with radius $r$ in all of the mass-redshift bins. This does
not come as a surprise since the inflowing mass should be roughly conserved.
The following systematic variations of the amount of inflow as a function of
host halo mass $M_{\rm vir}$ and redshift $z$ become apparent: (a) The averaged
amount of inflow increases with increasing host halo mass. (b) Both, the amount
of inflow as well as the halo mass increase by roughly the same factor (one
order of magnitude higher halo mass results in roughly one order of magnitude
higher inflow) and (c) the total amount of inflow decreases slightly with
increasing time (decreasing redshift). These trends are most convincingly seen
in gas.

In the following we will make frequent use of basic statistical measures of our
samples, namely the linearly averaged mean $\langle \dot{M} \rangle$, the mean
of the logarithm $(\mu_0)$ and the standard deviation $(\sigma_0)$. The reader
should note that those (unlike the later defined  $\mu_1$, $\mu_2$, $\sigma_1$,
$\sigma_2$ or $a_{1 / 2}$) are intrinsic properties of the distributions, that
are independent from any model, equation or fits to it. Their definitions are:
\begin{eqnarray}
\langle \dot{M} \rangle & = & {1 \over n_{\rm s}} \sum_{\rm i = 1}^{\rm n_{\rm s}}
\dot{M}_{\rm i} \nonumber \\
\mu_0    & = & {1 \over n_{\rm s}} \sum_{\rm i = 1}^{\rm n_{\rm s}} \log_{10}
\left(\dot{M}_{\rm i}\right) \nonumber \\
\sigma_0   & = & \sqrt{{1 \over n_{\rm s}} \sum_{\rm i = 1}^{\rm n_{\rm s}}
\left[ \log_{10} \left(\dot{M}_{\rm i}\right) - \mu_0 \right]^2}
\label{eqn:moments}
\end{eqnarray}
$n_{\rm s}$ is the total number of shells in the sample, so the number of
galaxies in the respective bin times the number of different radii which are
used. $\dot{M}_{\rm i}$ is the amount of gas inflow through the respective shell.

\begin{figure*}
\begin{center}
\psfrag{gas}[B][B][1][0]{gas}
\psfrag{$0$}[B][B][1][0]{$0$}
\psfrag{$0.5$}[B][B][1][0]{$0.5$}
\psfrag{$1$}[B][B][1][0]{$1$}
\psfrag{$1.5$}[B][B][1][0]{$1.5$}
\psfrag{$2$}[B][B][1][0]{$2$}
\psfrag{$2.5$}[B][B][1][0]{$2.5$}
\psfrag{$3$}[B][B][1][0]{$3$}
\psfrag{$3.5$}[B][B][1][0]{$3.5$}
\psfrag{$4$}[B][B][1][0]{$4$}
\psfrag{MN z = 1.57}[Bl][Bl][1][0]{{\MN} $z = 1.57$}
\psfrag{CDB z = 2.25}[Bl][Bl][1][0]{{\CD} $z = 2.25$}
\psfrag{ARP z = 1.14}[Bl][Bl][1][0]{{\C14} $z = 1.14$}
\psfrag{M = 1e12 MO}[Bl][Bl][1][0]{$M_{\rm vir} = 10^{12}$ M$_\odot$}
\psfrag{M = 8.6e11 MO}[Bl][Bl][1][0]{$M_{\rm vir} = 8.6 \times 10^{11}$ M$_\odot$}
\psfrag{M = 7.1e11 MO}[Bl][Bl][1][0]{$M_{\rm vir} = 7.1 \times 10^{11}$ M$_\odot$}
\psfrag{mu = 1.72, s = 0.22}[Bl][Bl][1][0]{\textcolor{blue}{$\mu_0 = 1.72 \ \ \
\ \sigma_0 = 0.22$}}
\psfrag{mu = 1.90, s = 0.25}[Bl][Bl][1][0]{\textcolor{blue}{$\mu_0 = 1.90 \ \ \
\ \sigma_0 = 0.25$}}
\psfrag{mu = 1.82, s = 0.18}[Bl][Bl][1][0]{\textcolor{blue}{$\mu_0 = 1.82 \ \ \
\ \sigma_0 = 0.18$}}
\psfrag{mu1 = 1.67, s1 = 0.12}[Bl][Bl][1][0]{\textcolor{Gray}{$\mu_1 = 1.67 \ \
\ \ \sigma_1 = 0.12$}}
\psfrag{mu2 = 2.07, s2 = 0.26}[Bl][Bl][1][0]{\textcolor{green}{$\mu_2 = 2.07 \ \
\ \ \sigma_2 = 0.26$}}
\psfrag{mu1 = 1.82, s1 = 0.18}[Bl][Bl][1][0]{\textcolor{Gray}{$\mu_1 = 1.82 \ \
\ \ \sigma_1 = 0.18$}}
\psfrag{mu2 = 2.14, s2 = 0.30}[Bl][Bl][1][0]{\textcolor{green}{$\mu_2 = 2.14 \ \
\ \ \sigma_2 = 0.30$}}
\psfrag{mu1 = 1.79, s1 = 0.14}[Bl][Bl][1][0]{\textcolor{Gray}{$\mu_1 = 1.79 \ \
\ \ \sigma_1 = 0.14$}}
\psfrag{mu2 = 2.11, s2 = 0.22}[Bl][Bl][1][0]{\textcolor{green}{$\mu_2 = 2.11 \ \
\ \ \sigma_2 = 0.22$}}
\psfrag{a = 0.84}[Bl][Bl][1][0]{\textcolor{red}{$a_{1 / 2} = 0.84$}}
\psfrag{a = 0.76}[Bl][Bl][1][0]{\textcolor{red}{$a_{1 / 2} = 0.76$}}
\psfrag{a = 0.89}[Bl][Bl][1][0]{\textcolor{red}{$a_{1 / 2} = 0.89$}}
\psfrag{dP / d(log (dM / dt))}[B][B][1][0]
{${\rm d}P / {\rm d}({\rm log}\, \dot{M})$}
\psfrag{log10}[B][B][1][0]
{${\rm log}_{10} (\dot{M} \ [{\rm M}_\odot \ {\rm yr}^{-1}])$}
\includegraphics[width=17.73cm]{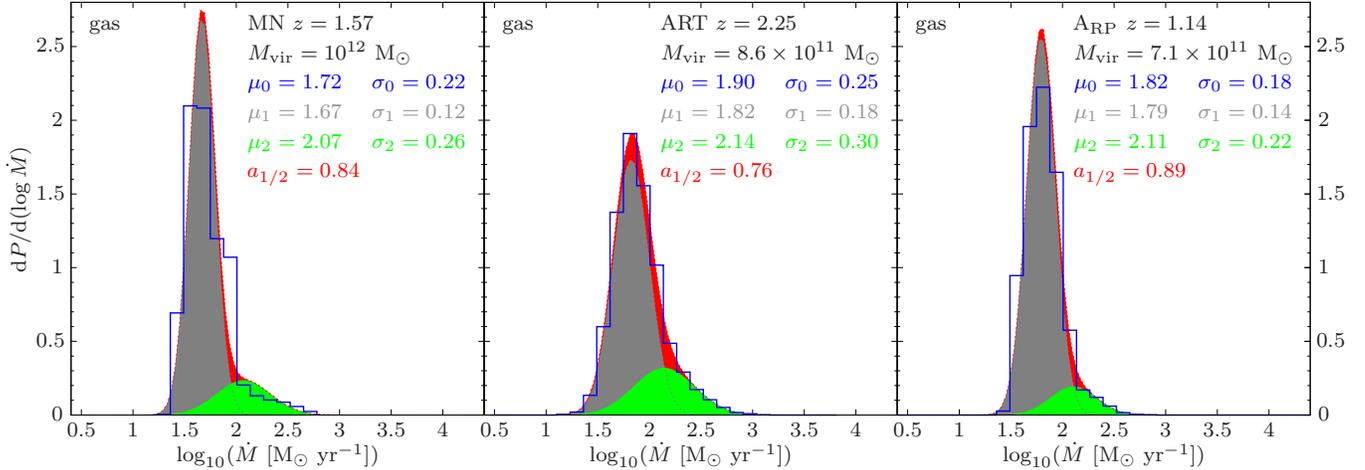}
\end{center}
\caption{Normalised distributions of the gas inflow of selected mass-redshift
bins for {\MN} (left panel), {\CD} (centre panel) and {\C14} (right panel) in
the radius range from $0.4 - 1.7 \ r_{\rm vir}$. The solid blue histograms are
the real distributions from the simulations. Shown in red are the individual
double-Gaussian (equation \ref{eqn:dblegauss}) fits to that particular panel. A
double-Gaussian consists of the sum of a first Gaussian (plotted in grey) and a
second Gaussian (plotted in green). $\mu_0$ and $\sigma_0$ denote the
statistical measures of the real underlying distribution, whereas $\mu_1$,
$\mu_2$, $\sigma_1$, $\sigma_2$ and $a_{1 / 2}$ from equation
(\ref{eqn:dblegauss}) are the best fit parameters for the double-Gaussian.
In all of these panels there is again a very beautiful double-Gaussian
behaviour, the quality of the fits is quite striking in all three panels. It is
most impressive for the right panel, since the {\C14} suite is our technically
most advanced simulation.}
\label{fig:distri}
\end{figure*}

The theoretical prediction \citep{neistein08} of the average gas inflow rate
$\langle \dot M \rangle$ into a host halo as a function of virial mass
$M_{\rm vir}$ and redshift $z$ is given by
\begin{equation}
\langle \dot M \rangle = 100 \ {{\rm M}_\odot \over {\rm yr}} \ \left({1 + z
\over 3.5} \right)^{2.5} \left({M_{\rm vir} \over 10^{12} \ {\rm M}_\odot} 
\right)^{1.15}.
\label{eqn:neistein}
\end{equation}
To compare our results to this prediction, the magnitude of the inflow in the
simulations will be averaged over the whole radius range for all mass-redshift
bins. The resulting averaged inflow values $M_{\rm inflow}$ of the gas are shown
in figure \ref{fig:Minfl} as a function of halo mass $M_{\rm vir}$ and redshift
$z$. In order to match $z = 2.46$ or $M_{\rm vir} = 10^{12} \ {\rm M}_\odot$ in
figure \ref{fig:Minfl} the data points had to be re-scaled according to the
above equation (i.e. the data in the left panel was rescaled according to
$\langle \dot M \rangle \propto [1 + z]^{2.5}$ and the data in the right panel
was rescaled according to $\langle \dot M \rangle \propto [M_{\rm vir}]^{1.15}$).
The original values of redshift (left panel) or host halo mass (right panel)
are given in colour coding. Equation (\ref{eqn:neistein}) is overlain as solid
black lines in both panels.

The simulated values for the inflow of the {\MN} and the {\CD} simulations are
consistent with the theoretical predictions. The agreement is robust over three
orders of magnitude in halo mass and also over the entire redshift range. The
simulated inflow values of {\C14} on the other hand are higher than the other
two suites of simulations and the theoretical prediction. Interestingly the
{\C14} suite clearly follows the same slopes with respect to both, the redshift
and the halo mass, still. This parallel shift by a constant factor $f_{\rm
rec}(\dot{M}) = \fa$ is due to the fact that the {\C14} simulations have more
massive outflows and therefore more reaccretion due to recycling. Since
equation (\ref{eqn:neistein}) describes only the primary accretion via cold
streams and not the secondary accretion due to recycling, the {\C14}
simulations should be scaled down by this factor. We will also see a similar
effect of recycling later in the paper during the fitting procedure of the
distributions.

Accounting for recycling in the way described leads to an excellent agreement
between the theoretical prediction and all the simulations including the {\C14}
suite, too. Only the highest redshift / lowest mass bins of both the {\MN} and
the {\CD} suite of simulation lie a little bit further away from the relation.
This is most probably due to the fact that these two bins are in a point in
parameter space where the simulations have a lower resolution and might
therefore be more vulnerable to numerical effects.

Gas supply of high mass galaxies $(M_{\rm vir} > 10^{12}$ M$_\odot)$ is
theoretically predicted \citep{db06} to be quenched at low redshifts $(z < 2)$.
It would be nice to spot this effect in simulations. However this turns out to
be a difficult task to do: The {\CD} galaxies are all at roughly $M_{\rm vir}
\sim 10^{12}$ M$_\odot$, the {\C14} galaxies are even lighter and both are
therefore too light to test this prediction. Neither the $10^{12}$ M$_\odot$ nor
the $5 \times 10^{12}$ M$_\odot$ nor the $10^{13}$ M$_\odot$ bins of the {\MN}
simulations (table \ref{tab:bins}) show drops of the gas infall from $z = 2.46$
down to $z = 1.57$ (see the values presented in figure \ref{fig:Minfl}) that
exceed the statistical uncertainties (compare the 1-$\sigma$ standard
deviations shown in figure \ref{fig:avinflvsradcombi}). Unfortunately the
simulations do not produce galaxies in such high mass regimes at any earlier
epochs sufficiently frequent enough. More high resolution simulations of those
high sigma peak galaxies are needed to establish a possible quenching of gas
supply at low redshifts for high mass haloes in simulations. However such
simulations would be computationally extremely demanding.

\section{Distributions}
\label{sec:distri}

\begin{figure*}
\begin{center}
\psfrag{MvMO}[B][B][1][0] {$M_{\rm vir} \ [{\rm M}_\odot]$}
\psfrag{a}[Bl][Bl][1][0] {$a_{1 / 2}$}
\psfrag{s}[B][B][1][0] {$\sigma_1$, $\sigma_2$}
\psfrag{s1}[Bl][Bl][1][0] {$\sigma_1$}
\psfrag{s2}[Bl][Bl][1][0] {$\sigma_2$}
\psfrag{m}[B][B][1][0] {$\mu_1$, $\mu_2$}
\psfrag{m1}[Bl][Bl][1][0] {$\mu_1$}
\psfrag{m2}[Bl][Bl][1][0] {$\mu_2$}
\psfrag{unrescaled}[Bl][Bl][1][0] {unrescaled}
\psfrag{gas}[Bl][Bl][1][0] {gas}
\psfrag{MN}[Bl][Bl][1][0] {{\MN}}
\psfrag{CDB}[Bl][Bl][1][0] {{\CD}}
\psfrag{ARP}[Bl][Bl][1][0] {{\C14}}
\psfrag{z}[B][B][1][0] {$z$}
\psfrag{$0$}[B][B][1][0] {$0$}
\psfrag{$0.1$}[B][B][1][0] {$0.1$}
\psfrag{$0.2$}[B][B][1][0] {$0.2$}
\psfrag{$0.3$}[B][B][1][0] {$0.3$}
\psfrag{$0.4$}[B][B][1][0] {$0.4$}
\psfrag{$0.5$}[B][B][1][0] {$0.5$}
\psfrag{$0.6$}[B][B][1][0] {$0.6$}
\psfrag{$0.7$}[B][B][1][0] {$0.7$}
\psfrag{$0.8$}[B][B][1][0] {$0.8$}
\psfrag{$0.9$}[B][B][1][0] {$0.9$}
\psfrag{$1$}[B][B][1][0] {$1$}
\psfrag{$1.5$}[B][B][1][0] {$1.5$}
\psfrag{$2$}[B][B][1][0] {$2$}
\psfrag{$2.5$}[B][B][1][0] {$2.5$}
\psfrag{$3$}[B][B][1][0] {$3$}
\psfrag{$3.5$}[B][B][1][0] {$3.5$}
\psfrag{$4$}[B][B][1][0] {$4$}
\psfrag{$1e11$}[B][B][1][0]{$10^{11}$}
\psfrag{$1e12$}[B][B][1][0]{$10^{12}$}
\psfrag{$1e13$}[B][B][1][0]{$10^{13}$}
\includegraphics[width=17.73cm]{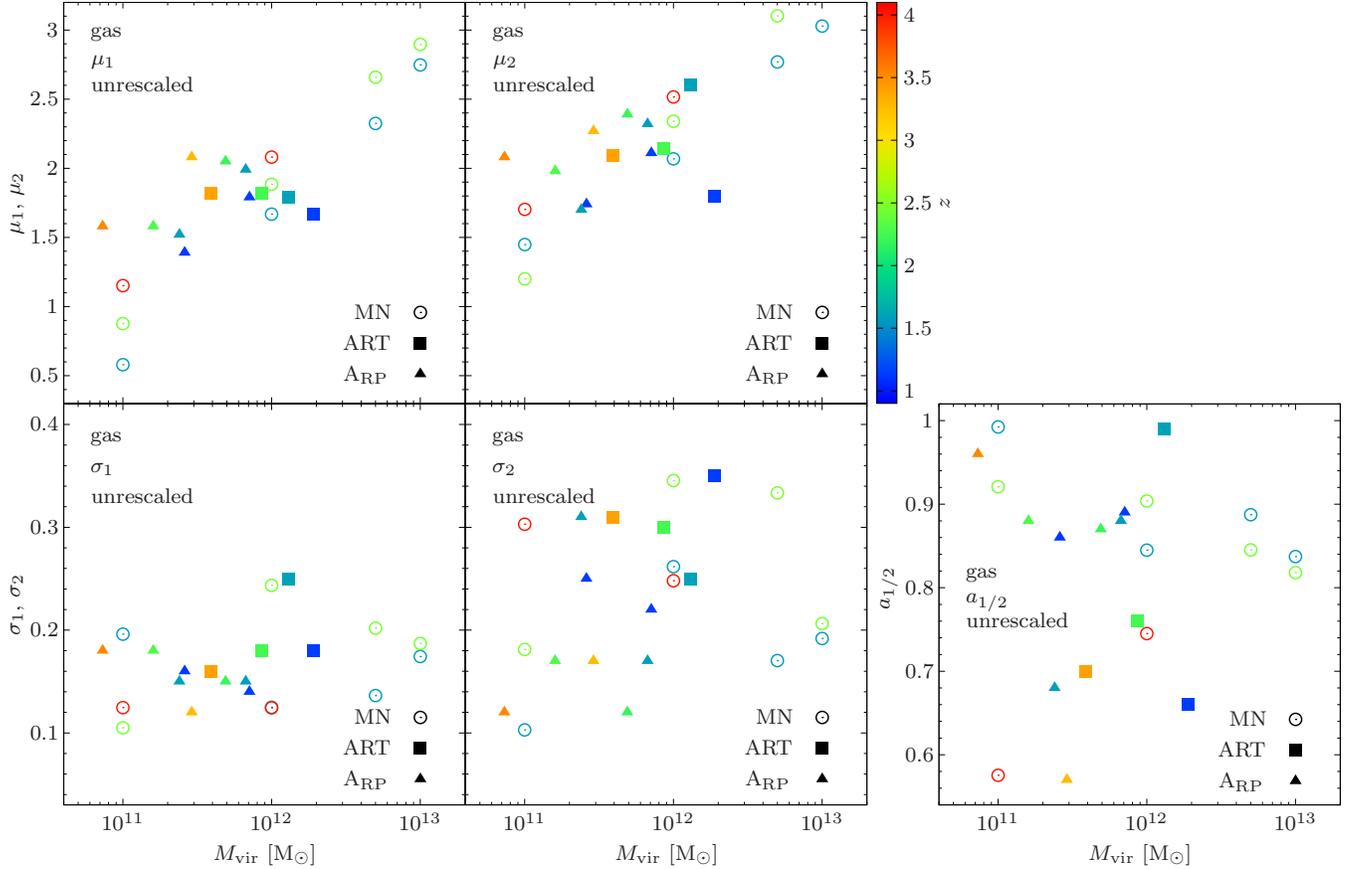}
\end{center}
\caption{Results of the fits from equation (\ref{eqn:dblegauss}) to the gas
distributions of all of the bins mentioned in table \ref{tab:bins}, examples
are plotted in figure \ref{fig:distri}. Shown are the variations of the five
parameters $\mu_1$, $\mu_2$, $\sigma_1$, $\sigma_2$ and $a_{1 / 2}$ from
equation (\ref{eqn:dblegauss}) as a function of halo mass $M_{\rm vir}$ and
redshift $z$. These are the parameters of a fit. In this plot the data is
completely unrescaled, unlike the same data shown in figure
\ref{fig:fitallparam}. All {\C14} values for $\mu_1$ and $\mu_2$ are due to
effects of recycling higher than expected. Physically these parameters show
that (a) there is more absolute inflow into heavier haloes, (b) there is more
absolute inflow at earlier cosmic times and there are heavier mergers happening
(c) at earlier epochs and also for (d) higher mass host haloes. An analogue
plot for the baryons was also produced, since it looks very similar to this one
it was omitted.}
\label{fig:param}
\end{figure*}

In section \ref{sec:raw}, the average amount of inflow was presented. In this
section, the behaviour of the underlying distributions is discussed as a
function of host halo mass and redshift. So the likelihoods of certain
deviations from the means presented in section \ref{sec:raw} are determined.
For this the amount of inflow is measured through thin spherical shells at a
number (200) of different radii between  $0.4 - 1.7 \ r_{\rm vir}$ around all
available galaxies separately. This is done for every redshift and mass bin
from all three suites of simulations. Having collected all these data no
average as in section \ref{sec:raw} is computed, but instead histograms how
often a certain inflow value occurs are compiled. All measurements of the
amount of inflow are treated equal, regardless of the radius they were obtained
at. The resulting normalised distributions of the gas inflow are plotted in
figure \ref{fig:distri} for all three suites of simulations ({\MN}, {\CD} as
well as {\C14}) as solid blue histograms. Similar figures showing the
distributions of the baryonic inflow instead of the gas were produced, found to
look extremely similar and therefore omitted. Some trends are seen, most
striking: the overall mean ($\mu_0$, as defined in equation \ref{eqn:moments})
of the sample distributions (examples are shown in figure \ref{fig:distri} as
solid blue histograms) appears to increase with increasing mass and also with
increasing redshift.

The best description for those histograms is the sum of a strong first Gaussian
(plotted in grey) with a smaller mean $(\mu_1)$ and an also smaller standard
deviation $(\sigma_1)$ and a weaker second Gaussian (plotted in green) having a
bigger mean $(\mu_2)$ and a bigger standard deviation $(\sigma_2)$. This sum
(plotted in red) can be expressed as the following ``double-Gaussian''
distribution:
\begin{eqnarray}
f(x) & = & {a_{1 / 2} \over \sigma_1 \sqrt{2 \pi}}\, {\rm exp}\left[{-(x -
\mu_1)^2 \over 2 \sigma_1^2}\right] \nonumber \\ & + & {1 - a_{1 / 2} \over
\sigma_2 \sqrt{2 \pi}} \, {\rm exp} \left[{-(x - \mu_2)^2 \over 2
\sigma_2^2}\right]
\label{eqn:dblegauss}
\end{eqnarray}
The variable $a_{1 / 2}$ gives the fractional strength of the first to the second
Gaussian. This model reflects the two different physical origins of
the inflow: The majority of the gas inflow is expected to flow into the host
halo by smooth accretion and a much smaller amount is expected to enter the
host by merger events. Each of these mechanisms is accounted for by one of the
two Gaussians of the distribution. Inflow by smooth accretion will naturally
occur with a lower amount of inflow at any one instant (small $\mu_1$) but
contribute much more to the inflow (big $a_{1/2}$) whereas inflow by merger
events will have a much higher amount of inflow at any one instant (big
$\mu_2$) but contributes much less to the inflow (small $1 - a_{1/2}$). So the
first Gaussian of this distribution represents the infall by smooth accretion
whereas the second Gaussian of this distribution represents the infall by
merger events.

\begin{figure*}
\begin{center}
\psfrag{MvMO}[B][B][1][0] {$M_{\rm vir} \ [{\rm M}_\odot]$}
\psfrag{mu2}[B][B][1][0]{$\mu_2$}
\psfrag{mu1}[B][B][1][0]{$\mu_1$}
\psfrag{gas, mu2}[Bl][Bl][1][0]{gas, $\mu_2$}
\psfrag{gas, mu1}[Bl][Bl][1][0]{gas, $\mu_1$}
\psfrag{baryons, mu2}[Bl][Bl][1][0]{baryons, $\mu_2$}
\psfrag{baryons, mu1}[Bl][Bl][1][0]{baryons, $\mu_1$}
\psfrag{MN}[Br][Br][1][0] {\MN}
\psfrag{CDB}[Br][Br][1][0] {\CD}
\psfrag{ARP}[Br][Br][1][0] {\C14}
\psfrag{ARP(mu1)}[Br][Br][1][0] {{\C14} scaled by $f_{\rm rec}(\mu_1)$}
\psfrag{ARP scaled by frec(mu2)}[Bl][Bl][1][0] {{\C14} scaled by
$f_{\rm rec}(\mu_2)$}
\psfrag{ARP scaled by frec(mu1)}[Bl][Bl][1][0] {{\C14} scaled by
$f_{\rm rec}(\mu_1)$}
\psfrag{z + 1}[B][B][1][0] {$z + 1$}
\psfrag{$1$}[B][B][1][0] {$1$}
\psfrag{$2$}[B][B][1][0] {$2$}
\psfrag{$2.5$}[B][B][1][0] {$2.5$}
\psfrag{$3$}[B][B][1][0] {$3$}
\psfrag{$3.5$}[B][B][1][0] {$3.5$}
\psfrag{$4$}[B][B][1][0] {$4$}
\psfrag{$4.5$}[B][B][1][0] {$4.5$}
\psfrag{$5$}[B][B][1][0] {$5$}
\psfrag{$1e11$}[B][B][1][0]{$10^{11}$}
\psfrag{$1e12$}[B][B][1][0]{$10^{12}$}
\psfrag{$1e13$}[B][B][1][0]{$10^{13}$}
\psfrag{scaled to 1e12 MO}[Bl][Bl][1][0]{scaled to $M_{\rm vir} = 10^{12}$
M$_\odot$}
\psfrag{scaled to z = 2.46}[Bl][Bl][1][0]{scaled to $z = 2.46$}
\psfrag{frec(mu1) = 1.45}[Bl][Bl][1][0]{$f_{\rm rec}(\mu_1) = 1.45$}
\psfrag{frec(mu2) = 1.20}[Bl][Bl][1][0]{$f_{\rm rec}(\mu_2) = 1.20$}
\includegraphics[width=13.00cm]{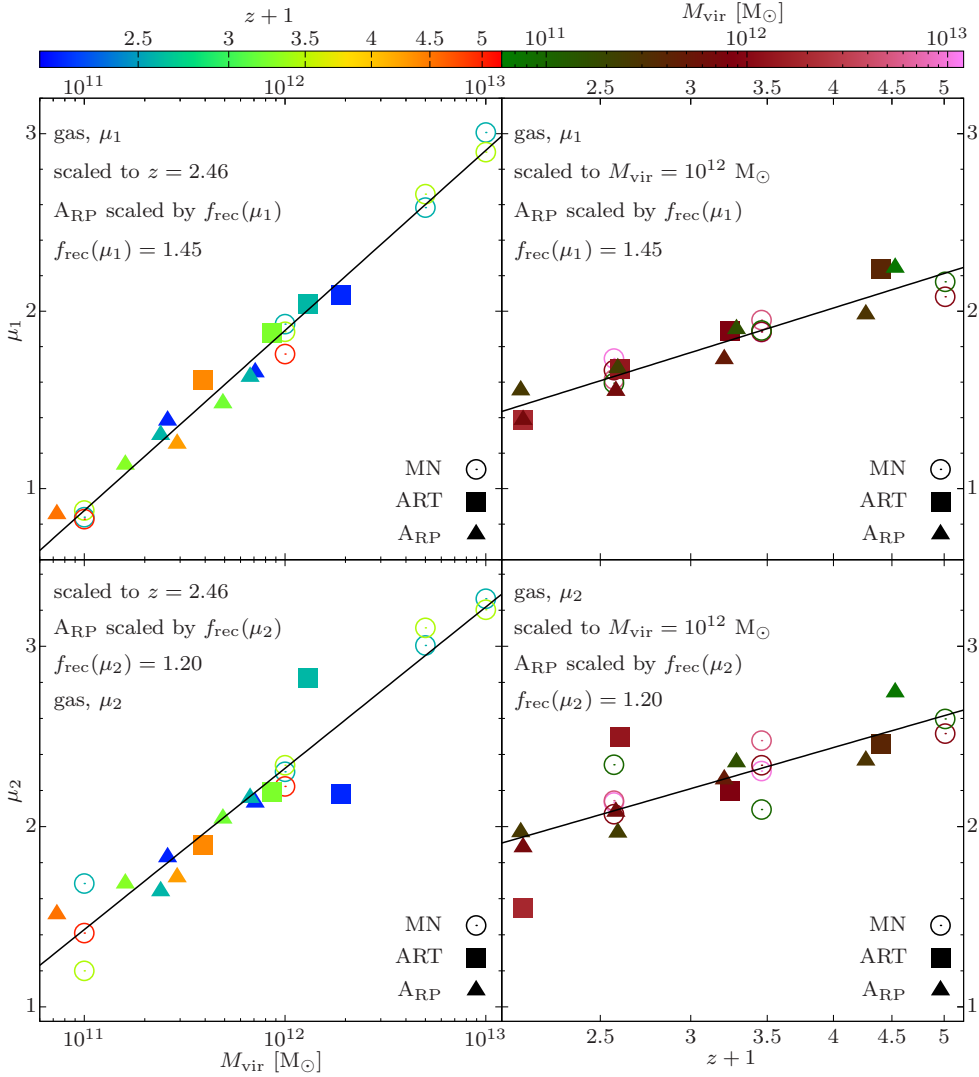}
\end{center}
\caption{Shown are the positions of the centres of the first ($\mu_1$, upper
panels) and the second ($\mu_2$, lower panels) Gaussians of the distributions
of the inflow as a function of mass (left panels) or as a function of redshift
(right panels). There are very strong logarithmic dependencies present in all
panels as described by equation (\ref{eqn:distriparam}) which is overlain as
solid black lines. The data points within the left panels were scaled with the
help of the scaling relation in the corresponding right panel to the values
indicated. The colour bar axes indicate the values the data points used to have
before rescaling. Please refer to figure \ref{fig:param} for the unrescaled
data. The panels show that our simple phenomenological model (equation
\ref{eqn:distriparam} and table \ref{tab:distriparam}) is an excellent
description of the data.}
\label{fig:fitallparam}
\end{figure*}

It is useful to introduce the double Gaussian functional form at this point for
several reasons: first, because it is a neat and very accurate description of
the behaviour of the inflow distribution in hydrodynamical simulations.
Secondly it reflects nicely the physics of the bimodal nature of the inflowing
material (mergers vs. smooth inflow). Last but not least this functional form's
parameters teach us directly the mean accretion rate for the mergers ($\mu_2$)
or for the smooth component ($\mu_1$), their standard deviations ($\sigma_2$,
$\sigma_1$) and maybe most importantly the relative strength of accretion via
the smooth component compared to the accretion via merger events ($a_{\rm 1 /
2}$). The above formula has already been used to describe the shape of the
specific star formation rate distribution at redshifts $z \le 2$ in
observational work \citep[][their equation 1]{sargent}. The similarity
indicates that smooth accretion might fuel main-sequence star formation on the
one hand and accretion through merger events might fuel star formation in
starburst events on the other hand.

The overall equation (shown in red) is an excellent description for all three
suites of simulations. Only exceptions are the very low mass  $(M_{\rm vir} =
10^{11}$ M$_\odot)$ {\MN} galaxies which are noisy and the low redshift $(z \le
1.7)$ {\CD} cases in which either the second Gaussian (green) is tiny or its
mean $(\mu_2)$ is not much bigger than the mean $(\mu_1)$ of the first Gaussian
(grey). The double-Gaussian behaviour can most beautifully been seen in the
technically most advanced {\C14} suite of simulations. Here all bins show an
almost prototypical behaviour in that sense.

We now want to look at the characteristics of the model. Therefore we fit the
distributions (examples are plotted as solid blue histograms in figure
\ref{fig:distri}) with equation (\ref{eqn:dblegauss}) using $\mu_1$, $\mu_2$,
$\sigma_1$, $\sigma_2$ and $a_{1 / 2}$ as free parameters. We do this for all
mass-redshift bins given in table \ref{tab:bins}. At this point we are seeking
the best fit for each bin individually. Examples of the resulting fits are
overplotted in figure \ref{fig:distri} as solid red areas. The values of the
parameters for all of those fits are presented in figure \ref{fig:param} as a
function of halo mass $M_{\rm vir}$ and redshift $z$. An analogue plot for the
baryons was also produced, since it looks very similar to the gas plot it was
omitted. The plots show that $\mu_1$ as well as $\mu_2$ increase with
increasing redshift, they also increase with increasing halo mass. A physical
interpretation of these trends is that there is more inflow into heavier haloes
and that there is also more inflow at earlier cosmic times. The {\C14} values
for $\mu_1$ and $\mu_2$ are considerably higher than their {\CD} or {\MN}
counterparts. This is an effect of recycling. For the $\sigma_1$ or $\sigma_2$
values the data are noisy and no clear trends are visible. Some of the values
for $a_{1 / 2}$ are also noisy. For gas however one can say that in more than
three quarter of all the cases $(17 / 22)$ $a_{1 / 2}$ is above 0.75. The mean
value for $a_{1 / 2}$ averaged over all 22 mass-redshift bins from the three
different suites is $<a_{1 / 2}> = 0.82$. These two numbers indicate according
to the physical interpretation of the double-Gaussian curve that at least three
quarter (possibly rather four fifth) of the inflowing material is coming in via
smooth streams and only less than one quarter (rather one fifth) is coming in
via merger events. This is in agreement with the findings of \citet{emilio} who
demonstrated that galaxies grow predominantly by smooth accretion from
cosmological filaments.

From the values shown in figure \ref{fig:param} we deduce that the mean
accretion rate for the mergers (given by $\mu_2$) is larger than that of the
smooth component (given by $\mu_1$). In log-space it is higher by $\sim 0.5$
which means that it is roughly a factor $2 - 3$ larger. These values indicate
in the terms of galaxy formation that significantly more material is coming
into the host halo via smooth accretion than via merger events. The standard
deviation of the merger accretion rate ($\sigma_2$) is $0.2 - 0.3$ dex, showing
no trend with mass or redshift. For the smooth component ($\sigma_1$) it is
$0.12 - 0.24$ dex. There are only marginal differences between gas and baryons.
The gas seems to account for the majority of the mass of the baryons. We
calculated the gas fraction from the simulations and find that $f_{\rm gas} =
85\,\%$. This is the case for all three suites of simulations, {\MN}, {\CD} as
well as {\C14}.

We want to have a more general model and found that the behaviour with mass and
redshift of two out of the five parameters, namely of $\mu_1$ and of $\mu_2$
can best be described by a logarithmic relation:
\begin{equation}
{\bf X} = {\bf A_X} \ \ln \left(z + 1 \right) + {\bf B_X} \ \ln \left({M_{\rm
vir} \over {\rm M}_\odot} \right) + {\bf C_X}
\label{eqn:distriparam}
\end{equation}
It is valid for the two parameters $\mu_1$ and $\mu_2$, denoted by ${\bf X}$,
whereas ${\bf A_X}$, ${\bf B_X}$ and ${\bf C_X}$ are the three free parameter
describing the logarithmic relation for both parameters. When fitting this
equation we had to account for the very strong signs of recycling found in the
{\C14} suite of simulations, as we already did for figure \ref{fig:Minfl}:
There we showed the amount of inflow and found that the values for the {\C14}
simulations had to be scaled down by some factor due to the effects of
rescaling. In the following we are going to fit two of the parameters of
equation (\ref{eqn:dblegauss}) which are also prone to be affected by
recycling. The positions of the centres of the first and the second Gaussian of
the distribution on the inflow ($\mu_1$ and $\mu_2$ as defined by equation
\ref{eqn:dblegauss}) in the {\C14} simulations are considerably higher than in
the {\MN} or the {\CD} simulations. However the reader should note that those
are completely different physical values than the average amount of inflow
$<\dot{M}>$ which are related but their relation is non-linear and therefore
the effect of recycling has a differently strong effect on each of them. I.e.
we have to scale the values for $<\dot{M}>$, $\mu_1$ and $\mu_2$ for the {\C14}
simulations by independent scaling factors. The $\mu_1$ values of the {\C14}
simulation are a factor of $f_{\rm rec}(\mu_1) = 1.45$ higher than the
corresponding values for $\mu_1$ of the {\MN} or the {\CD} simulations. The
$\mu_2$ values of the {\C14} simulation are a factor of $f_{\rm rec}(\mu_2) =
1.20$ higher than the corresponding values for $\mu_2$ of the {\MN} or the
{\CD} simulations. Again both factors remain constant over the whole redshift
and host halo mass range considered. We present the best-fitting values for
${\bf A_X}$, ${\bf B_X}$, ${\bf C_X}$ and $f_{\rm rec}(\mu_{1 / 2})$ in table
\ref{tab:distriparam}.

\begin{table}
\begin{center}
\setlength{\arrayrulewidth}{0.5mm}
\begin{tabular}{ccccc}
\hline
${\bf X}$  & ${\bf A_X}$ & ${\bf B_X}$ & ${\bf C_X}$ & $f_{\rm rec}(\mu_{1/2})$\\
\hline
$\mu_{\rm 1 gas}$       & 0.87   & 0.44   & -11.4  & 1.45 \\
$\mu_{\rm 2 gas}$       & 0.79   & 0.39   & -9.4  & 1.20 \\
\hline
\end{tabular}
\end{center}
\caption{The best-fitting values for two ($\mu_1$ and $\mu_2$) out of the five
parameters of equation (\ref{eqn:dblegauss}) when fitted with our model
(equation \ref{eqn:distriparam}) to the data of all bins as given in table
\ref{tab:bins} as shown in in examples in figure \ref{fig:distri} (red solid
areas and labels). The other three parameters ($\sigma_1$, $\sigma_2$ and
$a_{\rm 1 / 2}$) were far too noisy to allow any meaningful fits. The column
$f_{\rm frec}(\mu_{\rm 1 / 2})$ indicates the factor the {\C14} suite of
simulations was scaled down by to account for its strong recycling. The actual
comparison of the model (equation \ref{eqn:distriparam}) using the tabulated
values and the simulations can be seen in figure \ref{fig:fitallparam}, the
accuracy is exceptional.}
\label{tab:distriparam}
\end{table}

In figure \ref{fig:fitallparam} we compare the general model of equation
(\ref{eqn:distriparam}) with the parameters of the double-Gaussians of the
simulated distributions. Show are the positions of the centres of the first
($\mu_1$, upper panels) and the second ($\mu_2$, lower panels) Gaussian of the
distributions of the inflow as a function of mass (left panels) and as a
function of redshift (right panels). There are very strong logarithmic
dependencies present in all panels as described by our general model which is
overlain as solid black lines. The data points within each panel were scaled
with the help of the scaling relation in the corresponding other panel to the
values indicated. The colour bar axes indicate the values the data points used
to have before rescaling. Please refer to figure \ref{fig:param} for the
unrescaled data. The panels show that our simple phenomenological model
(equation \ref{eqn:distriparam} and table \ref{tab:distriparam}) is an
excellent description of the data.

\section{Conclusions}
\label{sec:conc}
In this paper we looked at the inflow rates of accretion along streams from the
cosmic web into galaxies at high redshifts using three sets of \textsc{amr}
hydro-cosmological simulations. We calculated the amount of inflow as a
function of radius, host halo mass and redshift. We computed the distribution
of inflow rates and we found the following:

\begin{itemize}

\item The inflow rates are roughly constant with radius, their behaviour with
host halo mass and redshift follows clearly the predictions of
\citet{neistein08}.

\item The distributions of the log accretion rates can be very well described
by a ``double-Gaussian'' functional form (equation \ref{eqn:dblegauss}) that is
the sum of two Gaussians, the primary corresponding to ``smooth'' inflow and
the secondary to ``mergers''.

\item Most of the amount of inflow ($>$ 80\%) is entering the halo at low
inflow rates (i.e. smooth accretion) and only a small portion ($<$ 20\%) of the
inflow is coming in at high inflow rates (massive merger events).

\item Two out of the five parameters of the double-Gaussian function (namely
$\mu_1$ and $\mu_2$) have a strong redshift as well as host halo mass
dependence. In equation (\ref{eqn:distriparam}), table \ref{tab:distriparam}
and figure \ref{fig:fitallparam} we present a simple phenomenological model,
which describes the shape of the double-Gaussian distribution as a function of
mass and redshift.

\item The standard deviation of the total accretion rate is $0.2 - 0.3$ dex,
showing no trend with mass or redshift.

\item The suite of simulations that include strong feedback shows massive signs
of accretion due to recycling: i.e. for the {\C14} simulations we see (compared
to the {\MN} or {\CD} simulations that have only weak feedback) an increase in
the amount of inflow by a factor $\sim \fa$, independent of mass and redshift.
Due to the same effect an increase in $\mu_1$ of the order of $\sim$ 1.45 and
an increase in $\mu_2$ of the order of $\sim$ 1.2 (see table
\ref{tab:distriparam}) is also seen.

\item There are hardly any differences between the gas and the baryons. Gas
seems to account for the majority of the mass of the baryons anyway and
additionally the star formation seem to locally follow the gas.

\item There is more absolute inflow into heavier haloes compared to lighter
haloes also there is more absolute inflow at earlier cosmic times compared to
later cosmic times.

\end{itemize}

A double-Gaussian distribution has already been found by \citet{sargent}, who
looked at the distributions of specific star formation rates at fixed stellar
mass in observations. They interpreted the bimodality of this functional form
as contributions from either main-sequence star formation or star formation
during starburst activity. The analogy between their and our findings indicates
that connections might exist between smooth accretion and main-sequence star
formation on the one hand as well as accretion through merger events and
starburst activity star formation on the other hand. However the reader should
note that just having the same functional form is no proof.

In our most advanced simulations, the {\C14} simulations, which include the
strongest feedback recipes, we see strong evidence for the reaccretion of
formerly ejected material. This suite of simulations has more massive outflows
and part of the outflowing material rains back again to the central galaxy. An
effect coined ``recycling'' \citep{oppenheimer}. The inflow in the {\C14} suite
of simulations is enhanced  due to recycling compared to the {\MN} or the {\CD}
simulations. This effect provides a third distinct accretion mode along with
the ‘cold’ and ‘hot’ modes described by \citet{keresa} and \citet{db06}.

One potential limitation of our simulations may arise from the artificial
pressure floor imposed in order to properly resolve the Jeans mass. This may
have an effect on the temperature and density of the densest and coldest parts
of the streams, with potential implications on the estimated inflow rates.
Still modern hydrodynamical simulation codes are the best available tools for
recovering the stream properties. With 17-70 pc resolution, and with proper
cooling below $10^4$K, these simulations provide the most reliable description
of the cold streams so far.

In a forthcoming companion paper \citep{mich5} we are going to address the role
of mergers versus smooth flows by analysing the clumpiness of the gas streams.
We evaluate each clump mass and estimate a mass ratio for the expected merger.
Finally we will look at the distribution of constituents amongst the clumps.

We conclude that gas is flowing into a galaxy's halo mainly as smooth
accretion flows with only a minority $(< 25\%)$ coming in via merger events
such as clumps or small satellite galaxies. The accretion rate distributions
can be described by a double-Gaussian decomposition (equation
\ref{eqn:dblegauss}). That this functional form is also used to describe the
specific star formation rate distributions in observations could indicate that
it is the smooth accretion that fuels main-sequence star formation on the one
hand and the accretion of material through merger events on the other hand that
fuels star formation of starburst events.

\section*{Acknowledgements}
Tobias Goerdt is a Lise Meitner fellow. We thank the anonymous referee for
useful comments which greatly aided the clarity of this work. We acknowledge
stimulating discussions with Nicolas Bouch{\'e}, Oliver Czoske, Natalia Romero
and Mark Sargent. Tobias Goerdt would like to thank the University Observatory
Munich where parts of this work where carried out for their hospitality. The
simulations were performed in the astro cluster at HU, at the National Energy
Research Scientific Computing Center (NERSC), Lawrence Berkeley National
Laboratory and at NASA Ames Research Center. Parts of the computational
calculations were done at the Vienna Scientific Cluster under project number
70522. This work was supported by FWF project number M 1590-N27 and by MINECO
project number AYA 2012-32295.

\bibliographystyle{mn2e}
\bibliography{accr20.bbl}

\label{lastpage}
\end{document}